# Portable device to determine particle asymmetry parameter

YULI W. HEINSON,[1*] CHRISTOPHER M. SORENSEN,[2,] AND RAJAN K. CHAKRABARTY[1*]

[1]*Department of Energy, Environmental and Chemical Engineering, Washington University in St. Louis, One Brookings Dr, St. Louis, MO 63132, USA*
[2]*Physics Department, Kansas State University, 2338 N. 17th St, Manhattan, KS 66506, USA*
*\*ywheinson@wustl.edu; chakrabarty@wustl.edu*

**Abstract:** Accurate characterization of the asymmetry parameter *g* is of crucial importance for many radiative transfer and climate models. Here, we present our portable light scattering (PLS) device which is designed for *in situ*, real-time, and contact-free measurements of the particles phase function with an integration time less than 1 second. With the PLS device, we have characterized carbonaceous aerosols, such as smoldering products from Alaskan peat, soot from a kerosene lamp and non-carbonaceous aerosols such as Arizona Road Dust. We found the *g* to be 0.664±0.002, 0.506±0.004, and 0.701±0.020 for those three kinds of particles, respectively. Compare to commercial nephelometers, our PLS device has reduced error in *g* due to angular truncation.

## 1. Introduction

The asymmetry parameter *g* is the weighted average of the normalized scattered intensity with the cosine of the scattering angle and thus describes the directionality of the scattered light from particles. It can be considered as the ratio of the light scattered in the forward direction over that of the backward direction. By definition, $1 \geq g \geq -1$. If the scattering is isotropic, $g = 0$. When $g > 0$, forward scattering dominates; when $g < 0$, backward scattering dominates. Being able to determine where the scattered light is directed is important for many systems; for example, how much incident solar radiation is scattered back into space by aerosol plumes. Thus, accurate values of *g* are essential in many radiative transfer and climate models to assess aerosol radiative effects of the climate [1, 2].

A mathematical description of light scattering from a particle or group of particles is simply given by the particle's 4 by 4 Mueller matrix operating on the incident beam's Stokes vector [3, 4]. The first element of the Mueller matrix is $S_{11}$, the phase function which represents the angular distribution of the scattered light intensity as a function of the scattering angle *θ*. It is normalized so that the integral of the scattered light is equal to 1. The *g* parameter [1, 2] is defined by integrating $S_{11}$ with the cosine operator over scattering angles

$$g = \frac{1}{2}\int_{-1}^{1} S_{11}(\theta)\cos\theta\, d(\cos\theta) \tag{1}$$

A common way to determine *g* in many experimental and field studies is by use of a nephelometer integrated with a backscatter shutter [5]. The nephelometer obtains the backscatter fraction *b* which is then related to *g* [6-8]. The relationship between *b* and *g* was studied and empirically found in [9]. However, the smallest angle a nephelometer can approach is typically ~7°, which could lead to a significant truncation error in b especially for large particles.

Unlike the methods stated above, we have developed our state-of-the-art portable light scattering (PLS) device with the angle detection range 0.7° to 162°. By assuming $S_{11}$ remains constant at $\theta \leq 0.7°$, which indeed is the case for particles with radius of gyration $R_g < 6.9$ μm, we determine the asymmetry parameter *g* by adapting the concept of Eq. (1) by integrating $S_{11}$, over angle range 0° to 162° for *in situ*, real-time, and contact-free measurements.

## 2. Instrument design

Figure 1 shows the schematic diagram of our PLS device which is designed to determine the asymmetry parameter *in situ* and in real-time. This whole device occupies a 15" x 15" 'shoebox' like area, which makes it highly portable. To measure $S_{11}$, the vertically polarized CW 532 nm incident laser beam passes through a 1/4 λ wave plate with its fast axis 45° from vertical to ensure the incident is light circularly polarized. The incident beam is directed to illuminate the aerosol sample coming out of a 1/4" inner diameter copper tubing. The scattered light is collected by two 512 channel detectors (Hamamatsu S3902-512Q). The scattered light from 0° to 9° is collected by a Fourier lens and the unscattered light (the incident laser beam) is focused by the Fourier lens to a waist. At the waist, a mirror with a 500 μm diameter through-hole oriented at 45° to the mirror normal allows the unscattered light to pass while the scattered light is reflected by the mirror. A small portion of the scattered light at small angles (0° ~ 0.7°) is lost through the mirror hole along with the unscattered light. The reflected scattered light is then collected by a lens to be projected onto the detector. The scattered light from 12° to 162° is collected by a custom elliptical mirror. The scattered light from 162° to 168° is omitted due to the weak signal and strong mirror edge effect. The intersection of the incident light and the aerosol is at the near focal point of the elliptical mirror; an iris with a 1.5 mm opening is placed at the far focal point where the scattered light is refocused and then thereafter the diverging scattered light is projected onto the second detector. Our device measures the scattered light at all angles simultaneously with the integration time less than one second. This device allows for quick and efficient procurement of data, eliminates problems regarding aerosol stability, and makes the detection at small angles easier. The two detectors are connected to a data acquisition box which feeds data into a computer. This design is adapted from the one described in [10-12] yet with a smaller dimension, finer angle resolution, and wider angle range leading to a precise determination of *g*.

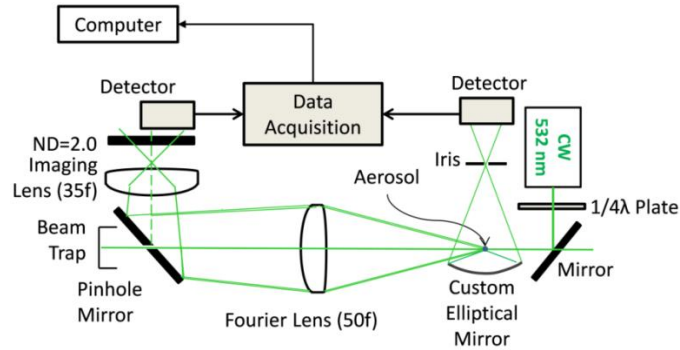

Fig. 1. A schematic diagram of the portable static light scattering (PLS) device. This device is designed to do *in situ*, contact-free, real-time measurements and it occupies a 15" x 15" area.

Since our device measures the relative scattered intensity $I(\theta)$ from 0.7° to 162°, assuming $I(\theta)$ remains constant at $\theta \leq 0.7°$, Eq. (1) then can be expressed as

$$g = \int_{-0.951}^{1} I(\theta) \cos\theta \, d(\cos\theta) \Big/ \int_{-0.951}^{1} I(\theta) \, d(\cos\theta) \qquad (2)$$

Plotting the scattered intensity vs. *q* where q=4π/λsin(θ/2), on a log-log scale, reveals Rayleigh regime, Guinier regime and power law regime [13-15]. From the Guinier regime, one can determine the particle radius of gyration $R_g$. Although this manuscript will skip the size analysis and focus on the asymmetry parameter $g$, we should address that the smallest detection angle of our PLS device limits the $R_g$ measurement to be less than 6.9 μm.

## 3. System test

Our device was tested with water droplets produced from a Collison 6-jet nebulizer (CH Technologies Inc., NJ, USA), similar as the calibration work done in [16]. Based on the size distribution of water droplets generated at 137.9 kPa (20 psi) [17] we calculated the theoretical scattered intensity to compare with the experimental data. Experimental data and theoretical calculations are in excellent agreement for both $\theta$ and $q$ space, as shown in Fig. 2A and B, respectively, except for the angle range 150° to 162°. This enhanced backscattering for the experiment compared to the theory has been observed in previous work [12] which has shown the phenomenon is a real effect and not due to calibration errors. Note that our smallest $q$ detection limit is $q=0.15$ μm$^{-1}$ (~0.7°) limited by the hole of the pinhole mirror. By integrating the scattering intensity, we found $g=0.818\pm0.005$ for these water droplets, as expected, which matched well with theory ($g=0.812$) generated in Mie theory [18].

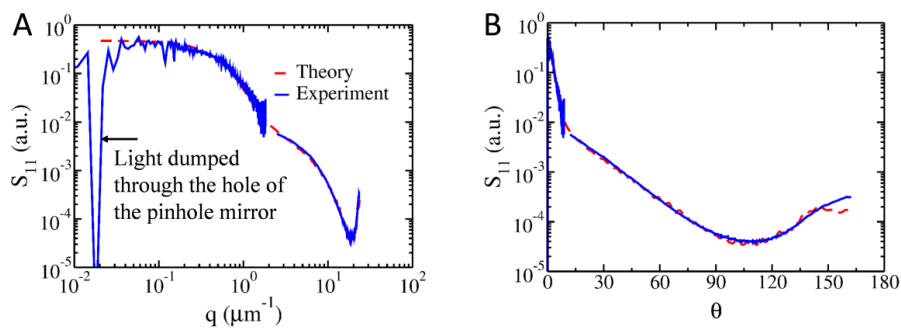

Fig. 2. Water droplets experiment for calibration. The experimental and theoretical phase functions match well for both (A) $q$ and (B) $\theta$ spaces. We found $g=0.818\pm0.005$ for these water droplets.

## 4. Carbonaceous and non-carbonaceous aerosol particles characterization

We have done some particle characterizations with this device for both carbonaceous and non-carbonaceous particles. For carbonaceous particle characterization, 10g of Alaskan peat was burned under smoldering condition in a 21 m$^3$ stainless steel chamber, similar as [19]. Smoldering peat fires produce spherical carbonaceous particles that absorb light in the near-UV which are commonly called brown carbon (BrC). One hour after the burn, the aerosol sample was collected in an 'iron lung'. The iron lung is made of a 16-gallon static-control flexible drum liner (McMaster-Carr 9772T43) inserted in a 10-gallon steel drum (McMaster-Carr 4115T12), with a sample port located on the lid. Aerosol samples can be inhaled and exhaled through the sample port by adjusting the pressure to the pressure port. Two plastic windows are mounted on the side of the drum at 90° from each other so one can observe the inhaling and exhaling process through one window by shining a flashlight through the other. The detail about the design of the iron lung is described in [20]. The exhaled aerosols were directed to the scattering volume of our PLS apparatus and measured $S_{11}$ was measured; the flow rate was set to be 0.6 L/min. At the same time, we also measured the size distribution of the BrC at the same flow rate with an SMPS. From previous work done in our lab, we have found the complex refractive index of BrC from Alaskan peat at $\lambda=532$ nm. Since in general they are spherical in morphology with small perturbations, using Mie theory, the refractive index, $m=1.45+0.002i$ [19], combined with the size distribution lead to the theoretical $S_{11}$. The experimental and theoretical $S_{11}$ match well, as shown in Fig. 3. The experimental curve only includes the side scattering data because for these small particles the scattering signal did not have a decent signal to noise ratio for the forward scattering. However, the side scattering begins to approach Rayleigh regime at small q, which is sufficient to determine

$g=0.664\pm0.002$, assuming the intensity remains constant at $\theta\leq12°$. The small uncertainty in $g$ is due to the use of the 'iron lung' which can significantly stabilize the aerosol flow and produce highly repeatable $g$ value. Compare to the theoretical $g=0.702$, the smaller value of the experimental $g$ is due to the enhanced backscattering of the measurement, as shown in Fig. 3B. Nonetheless, the measured $g$ is still within 6% of the theory assuming the BrC aerosols are perfect spheres.

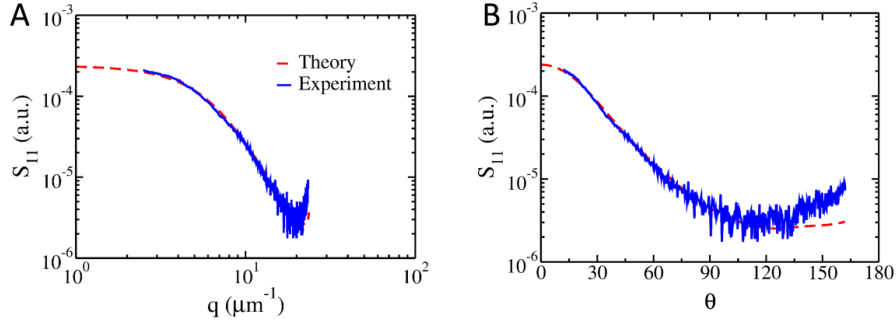

Fig. 3. The phase function of the smoldering product with Alaskan peat in (A) $q$ and (B) $\theta$ spaces. Again, the experimental and theoretical data match well before the enhanced backscattering takes place, with experimental $g=0.664\pm0.002$ and theoretical $g=0.702$. The flow rate was set to be 0.6 L/min.

Another carbonaceous aerosol we characterized with this device was the soot particles from a kerosene lamp. The experimental asymmetry parameter is $g=0.506\pm0.004$. The insets of Fig. 4 are the SEM images of the soot particles.

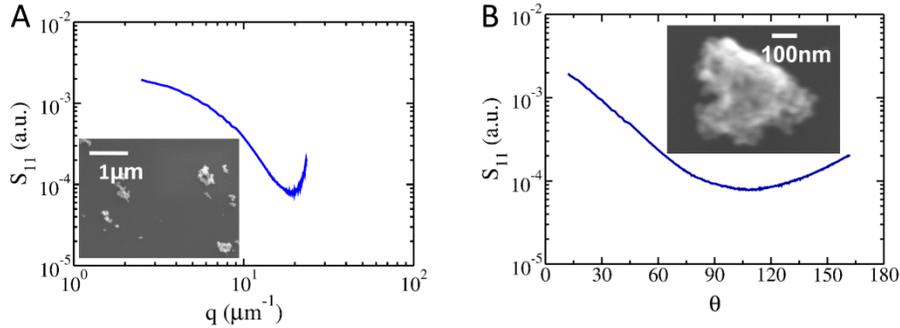

Fig. 4. The phase function of the soot particles from a kerosene lamp in (A) $q$ and (B) $\theta$ spaces with SEM images shown as the insets. The asymmetry parameter is $g=0.506\pm0.004$.

For non-carbonaceous particle characterization we purchased Ultrafine Arizona Road Dust (AZRD) from Powder Technology Inc. AZRD is a type of standard dust used for filter testing. It is also a used as an analogous model of the atmospheric dust particles. The dust particles were aerosolized in an aerosol generator. The design of the aerosol generator is describe in [21]. In our experiment, 5g AZRD was loaded into the aerosol generator chamber. The outgoing aerosolized particles were directed to the scattering volume. The phase function of AZRD is shown in Fig. 5. The kink at $q\approx20$ μm$^{-1}$ is expected as the SEM images (inset figures of Fig. 5) show AZRD has semi-spherical symmetry [22]. The experimental asymmetry parameter is $g=0.701\pm0.020$.

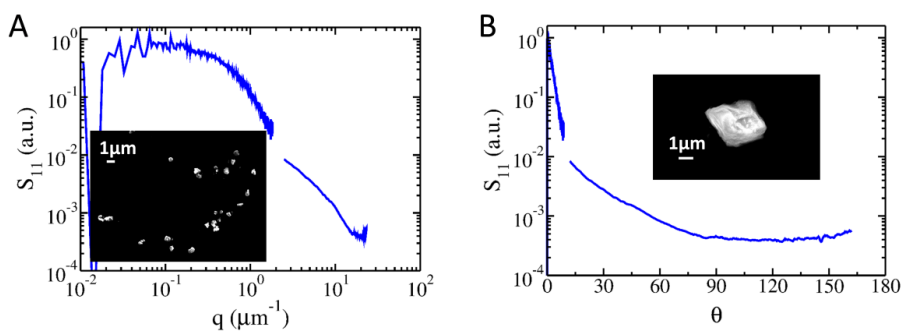

Fig. 5. The phase function of Ultrafine Arizona Road Dust (AZRD) in (A) $q$ and (B) $\theta$ spaces with SEM images shown as the insets. A kink occurs at $q \approx 20$ μm$^{-1}$ due to the semispherical symmetry shape. The asymmetry parameter is $g=0.701\pm0.020$.

## 5. Truncation error of *g*

Compared to a nephelometer, our smallest angle is one magnitude smaller, which would lead to much smaller truncation error of *g* when approaching large particles. We have compared the truncation error of *g* as a function of particle size. Using Mie theory, we ran spherical particles of geometric mean radius from 0.5 μm to 25 μm with the same geometric standard deviation of 1.2. Figure 6 shows the normalized radius probability as a function of geometric mean normalized radius (A) and the truncation error of *g* comparison between our PLS device and a nepholometer with forward truncation angle of 7° (B) as a function of geometric mean radius for two refractive indices m=1.33 for water and 1.95+0.79i for soot. Figure 6B clearly shows that our PLS device has much better performance with particles larger than the wavelength of the incident light (λ=532nm). The simple conclusion is that by having an order of magnitude smaller minimum angle of detection, one can accurately measure the asymmetry parameters for particles an order of magnitude larger.

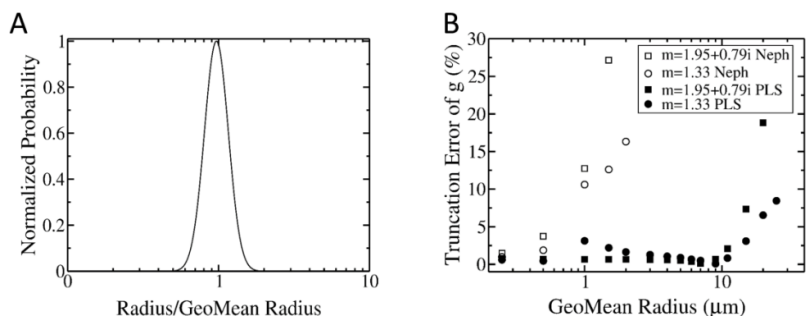

Fig. 6. (A) Normalized size distribution. (B) Truncation error of *g* as a function of geometric mean radius for our PLS device compare to a nephelometer for refractive indices m=1.33 and 1.95+0.79i.

## 6. Conclusion

We developed a portable light scattering device which is capable of determining the particle asymmetry parameter *g*. The device was calibrated and tested with water droplets. Using this device we have characterized carbonaceous particles such as brown carbon particles from Alaskan peat under a smoldering condition and soot particles from a kerosene lamp fire, and non-carbonaceous particles such as Ultrafine Arizona Road Dust. From the measurements we determined the *g* for the Alaskan peat brown carbon particles, kerosene lamp soot particles, and Ultrafine Arizona Road Dust are 0.664±0.002, 0.506±0.004, and 0.701±0.020, respectively. For particles larger than the wavelength of light the error of *g* due to angular truncation is greatly reduced compared to commercial nephelometers.

**Funding**

National Science Foundation (NSF) (AGS-1455215, CBET-1511964, AGM-1261651, AGS-1649783); National Aeronautics and Space Administration (NASA) (NNX15AI66G).

**Acknowledgements**

We would like to thank Dr. Huafang Li at the Institute of Materials Science & Engineering, Washington University in St. Louis for helping us collect the SEM images and Dr. William R. Heinson for his fruitful discussions on the theory prospective.